\documentclass[twocolumn,showpacs,preprintnumbers,amsmath,amssymb]{revtex4}
\usepackage{graphicx}% Include figure files
\usepackage{dcolumn}% Align table columns on decimal point
\usepackage{bm}% bold math

\begin{document}

%\preprint{APS/123-QED}
%
\newcommand{\bea}{\begin{eqnarray}}
\newcommand{\eea}{\end{eqnarray}}
\newcommand{\be}{\begin{equation}}
\newcommand{\ee}{\end{equation}}
%
%%%%%%%%%%%%%%%%%%  BOLDFACE GREEK LETTERS   %%%%%%%%%%%%%%%%
\newcommand{\xbf}[1]{\mbox{\boldmath $ #1 $}}
%%%%%%%%%%%%%%%%%%%%%%%%%%%%%%%%%%%%%%%%%%%%%%%%%%%%%%%%%%%%%

\title{Electromagnetic $N \to \Delta$ transition 
and neutron form factors~\footnote{published in Phys. Rev. Lett. {\bf 93}, 
212301 (2004).}} 

\author{A. J. Buchmann}
\affiliation{
Institute for Theoretical Physics \\ 
University of T\"ubingen \\ 
D-72076 T\"ubingen, Germany}
\email{alfons.buchmann@uni-tuebingen.de}

\date{\today}% It is always \today, today,

\begin{abstract}
%%%%%%%%%%%%%%%%%%%%%%%%%%%%%%%%%%%%%%%%%%%%%%%%%%%%%%%%%%%%%%%%

The $C2/M1$ ratio of the electromagnetic 
$N \to \Delta(1232)$ transition, which is important for 
determining the geometric shape of the nucleon, is shown to be 
related to the neutron elastic form factor ratio $G_C^n/G_M^n$.
The proposed relation holds with good accuracy for the entire
range of momentum transfers where data are available. 

\end{abstract}
%%%%%%%%%%%%%%%%%%%%%%%%%%%%%%%%%%%%%%%%%%%%%%%%%%%%%%%%%%%%%%%%%

% PACS, 
%the Physics and Astronomy Classification Scheme.
\maketitle

%%%%%%%%%%%%%%
%Introduction
%%%%%%%%%%%%%%
The regularities seen in the spectrum of excited states of a physical system
are usually due to an underlying symmetry. This is also the case in 
subnuclear physics, in particular in baryon physics. There, 
SU(3) flavor symmetry allows grouping the known baryons into 
singlets, octets, and decuplets~\cite{Gel64}.
Furthermore, SU(6) spin-flavor symmetry
unites the spin 1/2, flavor octet baryons ($2 \times 8$ states), 
among them the familiar proton and neutron, and the spin 3/2, 
flavor decuplet baryons ($4 \times 10$ states) into a common 
56-dimensional supermultiplet~\cite{Gur64,Beg64}.
These symmetries explain why the masses, 
electromagnetic moments, and other properties of baryons belonging 
to the same multiplet
follow a regular pattern. They arise mainly because octet and decuplet baryons 
are composed of the same spin 1/2, flavor triplet quarks merely coupled to 
different total spin and flavor.
% and because the quark-quark interaction
%has simple spin-flavor transformation properties.

\begin{figure}
\includegraphics{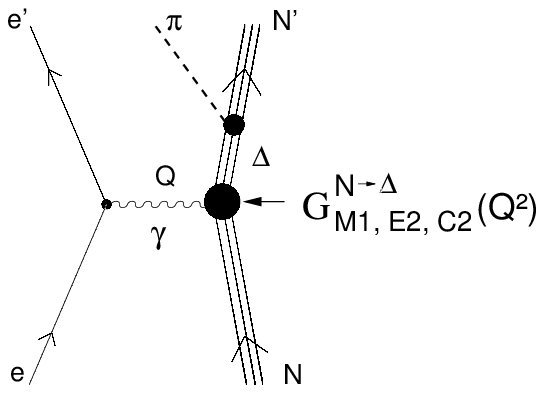}
\caption{\label{figure:scattering} 
The excitation of the $\Delta$ resonance by a virtual 
photon $\gamma$ of momentum $Q$ is described 
by the three electromagnetic transition form factors 
$G^{N \to \Delta}_{M1}(Q^2)$, $G^{N \to \Delta}_{E2}(Q^2)$, 
and $G^{N \to \Delta}_{C2}(Q^2)$. 
They can be determined by measuring 
the angular distribution of the decay pions in 
coincidence with the scattered electron. }
\end{figure}

The lowest mass member of the baryon flavor decuplet, called $\Delta(1232)$, 
with spin 3/2 and isospin 3/2 occupies a prominent place in baryon
spectroscopy not only because it has of all nucleon resonances 
the highest production cross section, but also because its properties 
are closely related to those of the nucleon. The $\Delta$ resonance is the 
lowest lying excited state of the nucleon $N(939)$ with the same quark content 
as the ground state. When produced in an electromagnetic process, such as 
electron-nucleon scattering (Fig.~\ref{figure:scattering}), 
parity invariance and angular momentum conservation 
restrict the $N \to \Delta$ excitation 
to magnetic dipole ($M1$), electric quadrupole ($E2$), and charge 
(or Coulomb) quadrupole ($C2$) transitions. 

At low momentum transfers the $N \to \Delta$ excitation is predominantly 
an $M1$ transition involving the spin and isospin flip of a single 
quark. The quadrupole amplitudes are only about $1/40$ of the 
dominant magnetic dipole amplitude.
Despite their smallnesss, the $C2$ and $E2$ multipoles have been the focus of 
many recent experimental~\cite{Ber03,Buu02,Joo02,Bar02} and
theoretical works~\cite{Idi04,Tia03,Jen02,Ale03,Hes02}. 
They are nonzero only if the geometric 
shape of the nucleon deviates from spherical symmetry~\cite{Hen01}.
From the corresponding quadrupole transition form factors
information on the spatial shape of the nucleon's charge distribution 
can be obtained. 

The purpose of this Letter is to show that the ratio of the 
$N \to \Delta$ charge quadrupole over magnetic dipole form factors, 
called the $C2/M1$ ratio, follows in good approximation 
the same curve as the ratio of the elastic neutron charge 
over magnetic form factors $G_{C}^n/G_{M}^n$ for the entire range
of momentum transfers where data are available.
This has not been noticed before.

Because the $N$ and $\Delta$ belong to the same 56-dimensional 
ground state multiplet of the SU(6) spin-flavor group their 
properties are related. In particular, the electromagnetic $N \to \Delta$ 
transition form factors are related to the electromagnetic elastic form 
factors of the nucleon~\cite{footnote0}.
This remains true even if the symmetry is broken. 
The experimentally observed breaking of SU(6) symmetry is not a 
fundamental objection against its usefulness.
If the relevant symmetry breaking mechanisms are included in the theory the 
resulting approximate symmetry leads to relations that are often very
well satisfied in nature~\cite{Leb95}. 

%%%%%%%%%%%%%%%%%%%%%%%%%%%%%%%%%%%%%%%%%%%%%%%%%%%%%%%%%%%%%%%%%%%%%%%%%
%             FIG. 1: Inelastic electron nucleon scattering
%%%%%%%%%%%%%%%%%%%%%%%%%%%%%%%%%%%%%%%%%%%%%%%%%%%%%%%%%%%%%%%%%%%%%%%%%
%%%%%%%%%%%%%%%%%%%%%%%%%%%%%%%%%%%%%%%%%%%%%%%%%%%%%%%%%%%%%%%%%%%%%

The SU(6) relation between the $N \to \Delta$ magnetic dipole transition 
form factor $G^{N\to \Delta}_{M1}(Q^2)$ and the elastic neutron magnetic 
form factor $G_M^n(Q^2)$ has been known for some 
time~\cite{Beg64}
\bea
\label{ffrel1}
G_{M1}^{N \to \Delta}(Q^2) &  = &  - \sqrt{2} \, \, G_M^n(Q^2).
\eea
Here, $Q$ is the four-momentum transfer of the virtual photon. 
At $Q^2=0$, both form factors are normalized to their 
magnetic dipole moments $\mu_{N \to \Delta}$ and $\mu_n$ 
\be
\label{ffrel1stat}
\mu_{N \to \Delta} = - \sqrt{2} \, \, \mu_n.
\ee
These relations also hold when second order SU(6) symmetry 
breaking operators are included~\cite{Leb95}, and have also been derived 
in the quark model with two-quark currents~\cite{Buc00,footnote2}.
They are violated only by three-quark currents~\cite{Dil00} or 
third order SU(6) symmetry breaking operators~\cite{Leb04}.
The latter are suppressed by a 
factor $1/N_c^2$ with respect to the leading term~\cite{Das95} so that these 
relations are valid in good approximation.

The other relation between the $N \to \Delta$ charge 
quadrupole transition form factor  $G^{N\to \Delta}_{C2}(Q^2)$ and the 
elastic neutron charge form factor $G_C^n(Q^2)$
\bea
\label{ffrel2}
G_{C2}^{N \to \Delta}(Q^2) &  = &  -\frac{3\,\sqrt{2}}{Q^2} G_C^n(Q^2) 
\eea
was unknown until quite recently~\cite{Buc00,footnote1}. 
If SU(6) symmetry were exact both $G_C^n(Q^2)$ and 
$G_{C2}^{N \to \Delta}(Q^2)$
would be zero. Spin-dependent 
two-quark terms in the charge density
break SU(6) symmetry~\cite{footnote3} and lead to nonzero form factors
which are related as in Eq.(\ref{ffrel2}).
 
In the $Q \to 0$ limit, Eq.(\ref{ffrel2}) reduces to 
a relation~\cite{Buc97} between the $N \to \Delta$ transition quadrupole
moment $Q_{N \to \Delta}$ and the neutron charge radius 
$r_n^2$
\be
\label{ffrel2stat}
Q_{N \to \Delta}= \frac{1}{\sqrt{2}} \, r_n^2,
\ee
which is in good agreement with recent extractions of 
$Q_{N \to \Delta}$ from the data~\cite{Tia03,Bla01}. 
This relation and its generalization to finite momentum transfers
in Eq.(\ref{ffrel2}) are of more general validity because they also
hold in a theory~\cite{Hes02}, which includes spin-dependent three-quark terms 
in the charge density, and for an arbitrary odd number of colors $N_c > 1$.
From Eq.(\ref{ffrel2stat}) we learn  that the small deviation of $r_n^2$ 
from zero and the deviation of the nucleon's geometric shape from spherical 
symmetry as manifested in a nonzero $Q_{N \to \Delta}$ are closely related 
aspects of nucleon structure. Both phenomena have their origin in a 
nonspherical cloud of quark-antiquark pairs in the 
nucleon~\cite{Hen01}. These pair degrees of freedom are effectively 
described by two- and three-quark currents~\cite{Buc91,Buc97}.

Experimental results are often given for the  $C2/M1$ ratio, which 
is defined in terms of the 
$N \to \Delta$ transition form factors times a kinematical 
factor~\cite{comment0,Jon73}
\be
\label{c2m1def}
    \frac{C2}{M1}(Q^2) 
= : \frac{\vert {\bf q} \vert \, M_N}{6} \, \,  
\frac{G_{C2}^{N \to \Delta}(Q^2)}{G_{M1}^{N \to \Delta}(Q^2)},
\ee
where $M_N$ is the nucleon mass and $\vert {\bf q} \vert $ is
the three-momentum transfer of the virtual photon in the 
$\gamma N$ center of mass frame~\cite{comment1}. 

Inserting the above form factor relations [Eq.(\ref{ffrel1}) 
and Eq.(\ref{ffrel2})], the $C2/M1$ ratio can be expressed as the 
product of $G_C^n/G_M^n$ and a factor
\be
\label{c2m1ratio}
    \frac{C2}{M1}(Q^2) =
\frac{\vert {\bf q} \vert}{Q} \,  \frac{M_N}{2 Q} \, \, 
\frac{G_{C}^n(Q^2)}{G_M^n(Q^2)} =: {\cal R}_n(Q^2). 
\ee 
We abbreviate this product as ${\cal R}_n(Q^2)$.
Thus, the inelastic $N \to \Delta$ and the elastic neutron form factor ratios 
are related.  The theoretical uncertainty of this relation
is mainly due to third order SU(6) symmetry breaking terms 
(three-quark currents) omitted in Eq.(\ref{ffrel1}). We estimate it
to be of order $1/N_c^2$ or 10$\%$ (slightly increasing the predicted
$C2/M1$ ratio).
 
To check whether Eq.(\ref{c2m1ratio}) is satisfied by the data,
we calculated the ratio ${\cal R}_n({\rm exp})$ 
using experimental results~\cite{Her99,Mad03} for $G_C^n/G_M^n$ in the 
range $Q^2=0-0.45$ GeV$^2$ and compared it with $C2/M1$ 
data~\cite{Bla01,Bec97,Got01,Sid71} 
from pionproduction experiments (see Table~\ref{tab:table1}). 
We found the agreement between both  
data sets to be astonishingly good~\cite{Gra01, comment3}. 
In particular, in the real photon limit $Q \to 0$ we obtained 
\be
\label{static}
\frac{C2}{M1}(0) \! =  \!
-\frac{M_{\Delta}^2 -M_N^2}{2 M_{\Delta}}\,
\frac{M_N}{12}\, \frac{r^2_n}{ \mu_n} = -0.031 
\ee
in good agreement with the experimental $E2/M1$ 
ratio obtained from pion-photoproduction
by different groups~\cite{Bla01,Bec97,comment2,Buc98}. 
This result explains the experimental value for the $C2/M1$ ratio
in terms of the charge radius and the magnetic moment of the neutron.
We understand therefore why $C2/M1(0)= -0.03$.

%%%%%%%%%%%%%%%%%%%%%%%%%%%%%%%%%%%%%%%%%%%%%%%%%%%%%%%%%%%%%%%%%%%%%%%%
%%%%%%%                   TABLE I
%%%%%%%%%%%%%%%%%%%%%%%%%%%%%%%%%%%%%%%%%%%%%%%%%%%%%%%%%%%%%%%%%%%%%%%%%%%%%%%%%%%%%%%%%%%%%%%%%%%%%%%%%%%%%%%
\begin{table}
\caption{\label{tab:table1} 
The ratio ${\cal R}_n({\rm exp})$ from the elastic neutron form factor data 
is compared with the ratio $C2/M1({\rm exp})$ extracted from 
pion-electroproduction experiments for $Q^2=0-4$ GeV$^2$.
A two-parameter fit of the experimental data for ${\cal R}_n$ 
using Eq.(\ref{Galster}) with $a=0.9$ and $d=2.8$~\cite{Gra01} 
is also listed.
}
\begin{ruledtabular}
\begin{tabular}{ccll}
$Q^2$    &  $ {\cal R}_n({\rm exp})$ & ${C2}/{M1}({\rm exp})$ & ${\cal R}_n$  
 \\ \hline
  0.00   &   -0.031(01)~\cite{Buc00} &  -0.030(03)~\cite{Bla01}  & -0.031 \\ 
         &                           &  -0.025(02)~\cite{Bec97}  &        \\  
  0.15   &   -0.050(11)~\cite{Her99} &  -0.055(04)~\cite{Got01}  & -0.047 \\
  0.29   &   -0.068(10)~\cite{Her99} &  -0.064(21)~\cite{Sid71}  & -0.054 \\
  0.45   &   -0.053(06)~\cite{Mad03} &  -0.075(15)~\cite{Sid71}  & -0.059 \\
  0.67   &   -0.059(12)~\cite{Roh99} &  -0.066(06)~\cite{Joo02}  & -0.064 \\
  1.13   &   -0.059(05)~\cite{Mad03} &  -0.079(09)~\cite{Joo02}   & -0.068 \\
  1.45   &   -0.077(07)~\cite{Mad03} &  -0.077(16)~\cite{Joo02}  & -0.069 \\
  1.80   &   -0.058{\phantom{(00)}}~\cite{Lom02} &  -0.116(31)~\cite{Joo02}    
                                                                 & -0.070 \\
 2.80    &   -0.061{\phantom{(00)}}~\cite{Lom02} &  -0.060(10)~\cite{Fro98}
                                                                 & -0.070 \\
 3.25    &   -0.066(30)~\cite{Lun93} &                            & -0.070 \\
 4.00    &   -0.078(43)~\cite{Lun93} &  -0.110(10)~\cite{Fro98}   
                                                                 & -0.069 \\
12.00   &              &                                         & -0.065 \\
$\infty$  &            &                                         & -0.061 \\
\end{tabular}
\end{ruledtabular}
\end{table}
%%%%%%%%%%%%%%%%%%%%%%%%%%%%%%%%%%%%%%%%%%%%%%%%%%%%%%%%%%%%%%%%%%%%%%%%%%%%%%%%%%%%%%%%%%%%%%%%%%%%%%%%%%%%%%%%

In the following, we will see that the range of validity of 
Eq.(\ref{c2m1ratio}) is not confined to low $Q^2$
but extends to the highest momentum transfers for which both ratios 
have been measured. In order to show that it is valid
at higher momentum transfers, I use recent $G_C^n/G_M^n$ data between 
$Q^2=0.45-1.45$ GeV$^2$ from double polarization experiments 
involving both electron and hadron spin polarization~\cite{Mad03,Roh99}, 
calculate ${\cal R}_n({\rm exp})$, and compare it with 
$C2/M1$ data~\cite{Joo02} at nearly the same momentum transfers 
(see Table~\ref{tab:table1}). 
Considering the experimental uncertainties of both experiments the 
agreement between ${\cal R}_n({\rm exp})$ and $C2/M1({\rm exp})$ is good.
 
At still higher momenta $Q^2=1.8-4.0$ GeV$^2$, I employ 
a recent fit to the experimental results for all four nucleon 
form factors~\cite{Lom02} and the SLAC data~\cite{Lun93} for the neutron 
elastic form factors, and calculate ${\cal R}_n({\rm exp})$. This is then 
compared with the electroproduction data $C2/M1({\rm exp})$~\cite{Joo02,Fro98}.
Table~\ref{tab:table1} shows that Eq.(\ref{c2m1ratio}) is satisfied 
within the experimental uncertainty.

In order to interpolate between experimental values and to extrapolate to 
higher $Q^2$, I also calculate the ratio ${\cal R}_n$ of Eq.(\ref{c2m1ratio})
(fourth column of Table~\ref{tab:table1})
using for the numerator a two-parameter fit~\cite{Gal71} of the $G_C^n$ data 
and for the denominator the dipole fit $G_D$ for $G_M^n$, i.e., 
\be 
\label{Galster}
G_C^n(Q^2)  =  -\mu_n \, \frac{ a \tau}{1 + d \tau} \, G_D(Q^2), \quad
G_M^n(Q^2)  =  \mu_n \, G_D(Q^2), \\
\ee
where $\tau = Q^2/(4 \, M_N^2)$ and 
$G_D= (1 + Q^2/\Lambda^2)^{-2}$ with $\Lambda^2=0.71$ GeV$^2$.
The $C2/M1$ ratio is then given in terms of the parameters $a$ and $d$,
which have been determined from the lowest moments of the experimental 
neutron charge form factor, namely the
neutron charge radius $r_n^2$, and the fourth moment $r_n^4$
(see Ref.~\cite{Gra01}).

%%%%%%%%%%%%%%%%%%%%%%%%%%%%%%%%%%%%%%%%%%%%%%%
%             FIG. 2: ratio C2/M1
%%%%%%%%%%%%%%%%%%%%%%%%%%%%%%%%%%%%%%%%%%%%%%%
\begin{figure}
\includegraphics{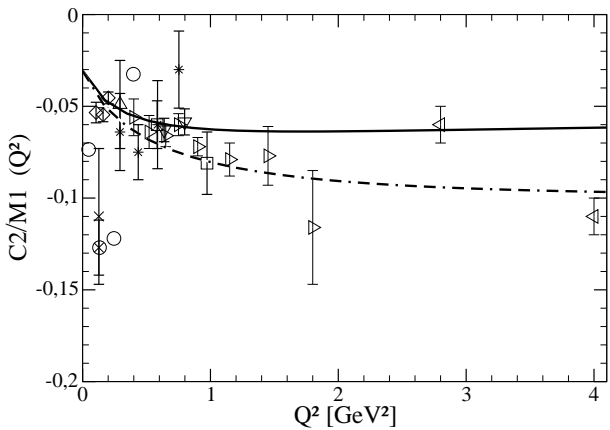}
\caption{\label{fig:fig2} 
The ratio ${\cal R}_n$ of Eq.(\ref{c2m1ratio}) 
calculated from a two-parameter fit of 
elastic neutron form factor data according to Eq.(\ref{Galster}).
Solid curve for $a=0.9$ and $d=2.8$, dashed-dotted curve for 
$a=0.9$ and $d=1.75$~\cite{Gra01}. This is compared with 
experimental results for the $C2/M1$ ratio extracted from
pion-electroproduction 
cross sections~\protect\cite{Joo02,Got01,Sid71,Fro98,Ald72}.
}
\end{figure}
%%%%%%%%%%%%%%%%%%%%%%%%%%%%%%%%%%%%%%%%%%%%%%%%%%%%%%%%%%%%%%%%%%%%%

In Fig.~\ref{fig:fig2} we plot ${\cal R}_n$ 
calculated from $G_C^n/G_M^n$ data using Eq.(\ref{Galster})
and compare with $C2/M1$ data from various pion-electroproduction 
experiments~\cite{Joo02,Got01,Sid71,Fro98,Ald72}. 
The solid and dashed-dotted lines correspond
to two different determinations of the parameter $d$.
Note the approximate constancy of the ratio ${\cal R}_n$ 
which is mirrored by the approximate constancy of the $C2/M1$ data
over a wide range of momentum transfers.
From Table~\ref{tab:table1} and Fig.~\ref{fig:fig2} we conclude that the 
equality of the inelastic and elastic form factor ratios predicted
by our Eq.(\ref{c2m1ratio}) is obeyed by the data for  
momentum transfers between 0 and 4 GeV$^2$.  
This means that the quark-antiquark degrees of freedom, 
which give rise to a nonzero $r_n^2$ and $Q_{N \to \Delta}$, 
also determine the corresponding form factors at higher $Q^2$.
It would be interesting to test the predicted 
constancy of this ratio at even higher momentum transfers. Work in this 
direction is in progress~\cite{Got04}.

Finally, we extrapolate our result to $Q^2 \to \infty$ and check whether 
${\cal R}_n(Q^2)$ is consistent with the 
perturbative QCD prediction for the asymptotic behavior of the $C2/M1(Q^2)$ 
ratio. From  Eq.(\ref{c2m1ratio}) I obtain using Eq.(\ref{Galster}) 
\be
\label{asymp}
{\cal R}_n(Q^2\to \infty)
=\frac{1}{4}\, \frac{M_N}{M_{\Delta}} \left (-\frac{a}{d}\right )\,= 
%=\frac{1}{4}\, \frac{M_N}{M_{\Delta}} \left (-\frac{r_n^2}{2 \mu_n}\, 
%\frac{2}{r_{N \to \Delta}^2-12/\Lambda^2}\right )\,= 
-0.061.
\ee
Thus, we see that the $C2/M1$ ratio asymptotically 
approaches a small negative constant determined by the neutron structure 
parameters $a$ and $d$. This is in qualitative agreement with expectations 
from perturbative QCD~\cite{Idi04} modulo logarithmic corrections.

Having gained some confidence in the validity of Eq.(\ref{ffrel2})
from low to high $Q^2$, we can Fourier transform it into 
coordinate space~\cite{footnote4}. The resulting quadrupole transition 
charge density $\rho_{C2}^{N \to \Delta}(r)$  might be useful for future
studies of the geometrical shape of the nucleon. 

In summary, recent measurements of the elastic neutron form factor 
ratio $G_C^n/G_M^n$ and the $C2/M1$ ratio in the electromagnetic 
$N \to \Delta$ transition show a remarkable agreement 
in sign and magnitude. This is true not only 
at $Q^2=0$ where C2/M1 is determined by the neutron charge
radius and magnetic moment 
but for the entire range of four-momentum transfers where data 
are available. In addition, the asymptotic $C2/M1$ ratio predicted 
on the basis of the $G_C^n/G_M^n$ ratio approaches a small negative 
constant in agreement with perturbative QCD. 

According to our theory, both ratios are related due to the 
underlying spin-flavor symmetry and its breaking by spin-dependent
two- and three-quark currents.  

The main conclusion of this paper is the observation that
the two data sets, which hitherto were thought to be quite independent of 
each other, satisfy the proposed relation Eq.(\ref{c2m1ratio}) within 
experimental uncertainties. This finding suggests that one can gain 
information concerning the geometric shape of the nucleon not only 
from the inelastic electron scattering cross section, but also from the 
elastic neutron form factor data. Conversely, one can determine the elastic 
neutron charge form factor from the $N \to \Delta$ charge quadrupole form 
factor extracted from pion-electroproduction data. 

%%%%%%%%%%%%%%%%%%%%%%%%%%%%%%%%%%%%%%%%%%%%%%%%%%%%%%%%%%%%%%%%%%%%%%%%%


\begin{thebibliography}{99}
\bibitem{footnote0}
The relations discussed in the following hold for both the 
$p \to \Delta^+$ and the $n \to \Delta^0$ transitions, which are 
generically denoted as $N \to \Delta$.
\bibitem{footnote2}
The general SU(6) and quark model derivations agree in the case of 
$\mu_{p \to \Delta^+}= -\sqrt{2} \mu_n$ because the latter depends only 
on the spin-flavor operators and wave 
functions dictated by SU(6) symmetry.
\bibitem{footnote3}
I conjecture that Eq.(\ref{ffrel2}) can also be derived 
in a group theoretical SU(6) tensor operator analysis.
\bibitem{footnote4} The Fourier transform of Eq.(\ref{ffrel2}) 
into coordinate space 
is (using Eq.(\ref{Galster}) and $m =: 2 M_N/\sqrt{d}$): 
$\rho_{C2}^{N \to \Delta}(r) = 
Q_{N \to \Delta} \,\frac{\Lambda^4 m^2}{4 \pi \, (\Lambda^2-m^2)^2} \,  
\biggl ( \frac{e^{- m r}}{r} - \frac{e^{- \Lambda r}}{r} +  
\frac{(m^2-\Lambda^2)}{2 \Lambda} \,  e^{- \Lambda r} \biggr ),$ with
$\int\! d^3 r \, \rho_{C2}^{N \to \Delta}(r) = Q_{N \to \Delta}$.
\end{thebibliography}
\end{document}